\documentclass{article}  

\usepackage{graphicx}

\begin{document}

\title{Multi-windowed defocused electronic speckle photographic system for tilt measurement}

\author{Jose M. Diazdelacruz\\
Department of Applied Physics, Faculty for Industrial Engineering, \\ Polytechnic University of Madrid.
\\  Jose Gutierrez Abascal 2. 28006 Madrid. Spain}

\begin{abstract}
Defocused speckle photography has long been used to measure rotations of rough surfaces. This paper explains how, by adding a suitably perforated mask, some measurement properties, such as range or lateral resolution, may be changed at wish. Particularly, the maximum measurable tilt can be significantly raised, although at the expense of poorer lateral resolution. Advantages over previously described techniques include independent tuning of speckle size and optical system aperture and more extended adaptability to different measuring needs. The benefits and disadvantages of the new and old techniques are thoroughly compared.
\end{abstract}

\maketitle %% NULL FUNCTION WITH LATEX 2e

\section{Introduction}

When a rough surface is illuminated by a laser beam, a granular effect is observed. This grainy pattern is known as speckle. Though this effect used to be considered a nuisance in early coherent photography, it has proved to be a valuable tool for the observation of mechanical transformations in solid bodies. Different optical systems can be used to obtain a record of speckles. Although surface roughness causes their formation, their statistics are generally quite independent of the particular microtopography of the surface and are more related to the optical set-up for illumination and recording.$^1$ Speckle patterns do not offer great information about the rough surface that generates it. However, the relation between the speckle patterns taken from a surface before and after a transformation may be used to obtain information about it. 

Assuming expanded beam illumination, object in-plane displacements (movements parallel to the surface under observation) may cause speckle pattern translations when the optical system is focused on the object surface, whereas out-of-plane rotations (tilts around in-plane axis) will not significantly affect the image. On the other hand, if the speckle pattern is recorded at the Fourier plane of the object surface, in-plane displacements will not appreciably change the speckles, although out-of-plane rotations will displace them.$^2$ When the optics is not focused on the object surface, the speckle pattern is said to be {\em defocused}. Defocused speckle photography is being used today as a mean to monitor and measure out-of-plane rotations in rigid solids or their distributions in the boundaries of elastic bodies. This paper develops an analysis of measuring characteristics such as lateral resolution or maximum measurable rotation in existing technologies, then presents a new enhancement for them and finally compares the measuring power in both systems.

\section{Antecedents}

The first paper describing a defocused two-exposure method to measure out-of-plane rotations was due to Tiziani,$^3$ and was later extended for vibration analysis.$^4$ The principle is simple and can be understood from the following consideration: when an initially vertical flat mirror undergoes an out-of-plane rotation around its horizontal direction, objects reflected by the mirror seem to be displaced upwards (or downwards) proportionally to their distance to the mirror. Considering the Kirchhoff approximation for the scattering from a rough body,$^{5,6}$ the surface can be thought of as a big set of small mirrors, slightly rotated from the mean plane. When the object is tilted, all those small mirrors experiment the same out-of-plane rotation and when focusing on a plane placed at a distance $\zeta$ from the surface, the amplitude distributions of the waves originated from all those micro-mirrors are all displaced by the same distance which is proportional to $\zeta$. The speckle pattern is thus displaced by an amount that depends on the tilt and can be used to measure it. Tiziani used collimated illumination and recorded the image at the back focal plane of the lens.

  \begin{figure}\centerline{{\includegraphics[width=10cm]{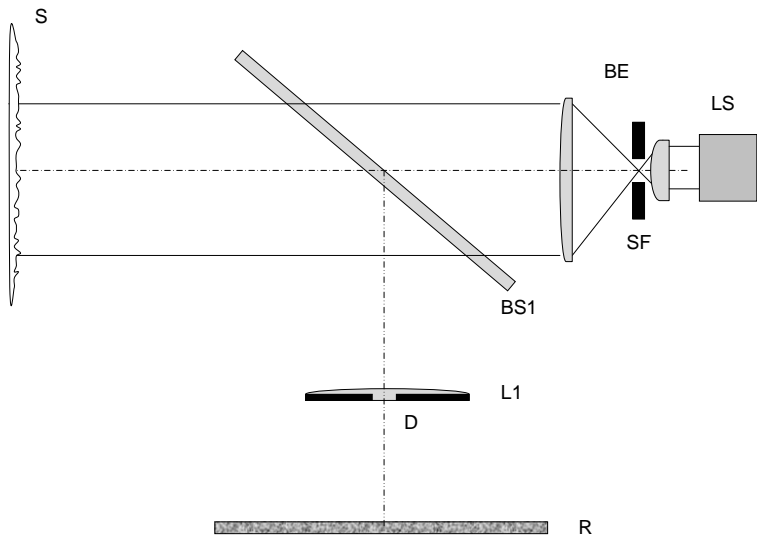}}}
  \caption{Experimental set-up using normal illumination and observation. }
  \end{figure}
  
If normal illumination and observation are used (Fig. 1), the speckle shift is given by$^7$
\begin{equation}
d_x= 2 f \beta
\end{equation}
\begin{equation}
d_y= - 2 f \alpha
\end{equation}
where $\alpha,\beta$ are the (small) rotation angles around the $x,y$ axis of a cartesian system placed on the mean plane of the object surface and $f$ is the focal length of the recording system. Lateral displacements do not appreciably alter these values.

Collimated illumination is difficult on extended objects. Gregory considered a divergent illumination and showed that when the optical system is focused on the plane than contains the image of the point source considering the object surface as a mirror, then the speckle shift does not depend on in-plane displacements, but only on out-of-plane tilts.$^{8\--10}$ Chiang and Juang  described a method to measure the change in slope by defocused systems.$^{11}$ A great number of later papers document the use of defocused speckle photography to measure in-plane and out-of-plane rotations and strains.$^{12\--15}$ This paper describes a new methodology that can be applied to almost all of them bringing advantages and disadvantages that will be discussed. 

After the double specklegram is obtained by addition (double exposure on the same photographic film) or subtraction (using a digital electronic camera), a pointwise analysis or a whole field filtering can be performed in order to know the lateral distribution of the displacement.$^{16,7}$  The first technique, illustrated in Fig. 2(a), uses a narrow laser beam that is made to go through the specklegram. The diffraction pattern is modulated by fringes whose directions are perpendicular to the speckle shift and whose spacing is
\begin{equation}
\sigma = \lambda \frac \xi {\sqrt{d_x^2 + d_y^2}}
\end{equation} 
where $\xi$ is the distance from the specklegram to the observation plane.

\begin{figure}\centerline{{\includegraphics[width=10cm]{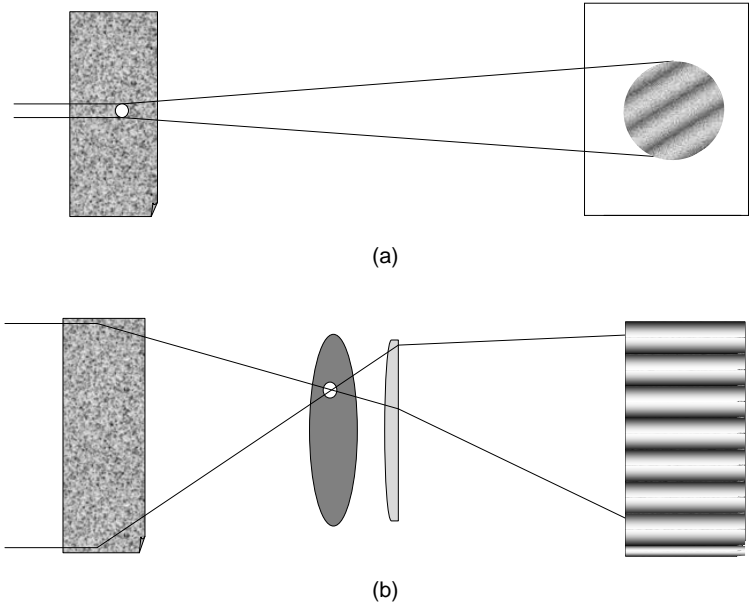}}}
  \caption{(a) Fringes arising from speckle shift around one point. (b) Fringes arising from whole field filtering.}
  \end{figure}

Whole field filtering, depicted in Fig. 2(b), uses a wide laser beam that goes through the specklegram and is filtered by a hole placed in a plane standing before a focusing lens that reconstructs an image of the specklegram. Only those areas whose modulated diffraction patterns do not have a dark fringe over the hole, will appear illuminated on the image plane. Thus, dark fringes in the detector can be mapped onto speckle shift values.

Today CCD cameras replace photographic films and processing of specklegrams is progressively performed through digital techniques,$^{17}$ although they usually simulate the analogic ones.

\section{Direct illumination system analysis}

We will consider the Tiziani optical system adapted for normal illumination represented in Fig.1. A laser source LS emits a light beam, that is expanded by BE and spatially filtered by SF. Half of the intensity goes through the beam splitter BS and reaches the rough surface S. Part of the light scattered by S is reflected by BS and recorded at the back focal plane R of the lens L1. We will refer to this set-up as DDSP(direct defocused speckle photography) set-up. The surface under observation is supposed to be a rough $L \times L$ square, placed at a distance $d$ from the objective lens of the optical system, measured along the optical axis (dashed in Fig. 1). The aperture diameter of the system is $D$, its focal length is $f$ and the sensing area is a $b \times b$ square whose resolution width is $\delta$. 

For a defocused recording system, the speckle size is given by$^{18}$
\begin{equation}
s=\frac{1.22 \lambda f}{D}
\end{equation}

When a part of the object surface undergoes an out-of-plane rotation $\gamma$, the light scattered from it experiments a rotation $2\gamma$ and completely falls off the aperture of the system when
\begin{equation}
2\gamma > \frac{D}{d}
\end{equation}
Thus  the maximum measurable rotation is
\begin{equation}
\Gamma = \frac{ f }{2d  F }
\end{equation}
being 
\begin{equation} F  = \frac{f}{D}\end{equation}
the f-number of the lens. Throughout this paper high apertures will mean low f-numbers and vice versa. In order to use the whole recording area, we should ensure that
\begin{equation}
\frac{L}{d} = \frac{b}{f}
\end{equation}
and accordingly
\begin{equation}
\Gamma=\frac{ b }{2L F }
\end{equation}

Neglecting diffraction effects (because we are using a defocused system), the diameter of the zone whose scattered light is incident on the same point at the detector plane is $D$, so the lateral resolution $\Delta$ of the system is equal to $D$

The relation
\begin{equation}
\frac{\Gamma}{\Delta} = \frac{b}{2f L}
\end{equation}
does not depend on the aperture.

By properly choosing the system aperture, the technique can improve the measuring range or the lateral resolution. Although, according to the equations, this versatility may seem feasible, speckle size considerations shrink the choice possibilities. Speckle size can not be smaller than the detector cell, because averaging destroys the speckle. On the other hand, there should be a large number of speckles, in order to accurately track their displacements, what sets an upper bound on speckle size. This limits the range of practical values for the aperture of the system. Often speckle size is made approximately equal to the detector cell size, so that
\begin{equation}
\delta = s =\frac{1.22 \lambda f}{D} \Rightarrow  F  = \frac{\delta}{1.22 \lambda}
\end{equation}
and therefore
\begin{equation}
\Gamma = \frac{1.22 b \lambda}{2L \delta}
\end{equation}
\begin{equation}
\Delta = \frac{1.22 f \lambda}{\delta}
\end{equation}

The main problem with the previously mentioned technique is its short range of measurements. The optical aperture of the system must remain small so that the speckle size is not smaller than the resolution of the CCD array. This fact limits the out-of-plane rotations whose speckle shifts can be recorded by the optical system. 

\section{Multi-windowed system analysis}

The DDSP set-up described in the previous section can be modified by the addition of a suitably perforated mask that enhances the adaptability of the system to a wider range of measuring needs. Fig. 3 represents an optical system where the laser source LS generates a beam which is expanded by a beam expander BE and spatially filtered through the pinhole SF. Then the collimated beam is split into multiple collimated beams by a multi-windowed mask MWM, which is shown in Fig. 4(a), with a rectangular array  of perforated circular holes of radius $a$ ($ \lambda \ll a $).  The beam splitter BS lets half of the light arrive at the object surface S and reflects light from S into the aperture of the lens L1. Finally a speckle pattern, as represented in Fig. 4(c), is recorded at the back focal plane R of the lens. The focusing and recording optics are implemented by an electronic CCD camera. We will refer to this set-up as MWDSP (multi-windowed defocused speckle photography) set-up. Other arrangements of holes are possible, although we will focus on the rectangular array depicted in Fig. 4(a), which is especially well suited for obtainig a square matrix $\gamma_{ij}=\gamma(i \Delta, j \Delta)$ of tilts, where $\Delta$ is the cartesian discretization step of the system. Equilateral triangular arrangements, as seen in Fig. 4(b), can also be used and are better than rectangular arrays when other parameters such as intensity efficiency or lateral resolution are concerned (this is because the non-perforated area and the distances between the centers of neighboring holes are smaller). Only the rectangular system will be considered here, because for other arrangements the analysis is practically identical. 

 \begin{figure}\centerline{{\includegraphics[width=10cm]{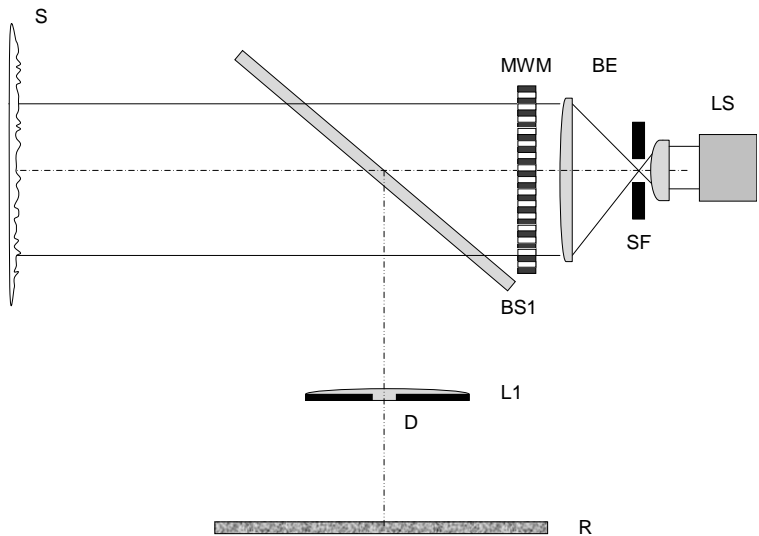}}}
  \caption{Multi-windowed normal illumination and observation set-up. }
  \end{figure}

We will now establish the main relations between relevant parameters of the system. In short, the number of illuminated spots on the object surface determines the lateral resolution of the system and the width of the detector area available for each, which then sets the maximum non-overlapping speckle displacement or, equivalently, the tilt range. 

If the object surface under observation and the sensing areas are $L \times L$ and $b \times b$ squares respectively, then, as in DDSP, the distance from the object to the camera, measured along the optical axis (dashed in Fig. 3), should be
\begin{equation}
d= \frac{L f}{b}
\end{equation}

In MWDSP, the average speckle size $s$ is given by$^{18}$
\begin{equation}
s=\frac {1.22 \lambda f}{2 a}
\end{equation}
where $a$ is the narrow beam radius, $\lambda$ is the laser wavelength and $f$ is the focal distance of the camera lens. 

Speckle size should be at least as wide as the detector cell, so that
\begin{equation}
 \delta \leq s \Rightarrow a \leq \frac{ 1.22 \lambda f}{2\delta} 
\end{equation}

 When the surface undergoes a tilt of angle $\gamma$, speckles are displaced by a distance $\ell$,$^7$ 
\begin{equation}
\ell= 2 \gamma f
\end{equation}
If we consider the maximum tilt to be $\Gamma$, then the maximum displacement will be
\begin{equation}
\ell_M = 2 \Gamma f
\end{equation}
Provided that the radius of the speckle circle is maximum, being $N$ the number  of rows, we have
\begin{equation}
N\leq \frac{b}{4 \Gamma f}
\end{equation}

The speckle circle radius $g$ is
\begin{equation}
g = \frac{f D}{d} = f  \frac{b}{2L F } 
\end{equation}
where $ F $ is the f-number of the camera lens. The non-overlapping condition for the recorded speckle circles is given by
\begin{equation}
g\leq \frac{b}{2N}
\end{equation}
so that
\begin{equation}
 F  \geq \frac{ fN}{ L }=\frac{b}{2 L \Gamma }
\end{equation}
and the usage of the detector area is maximized when
\begin{equation}
 F  =  \frac{ b}{ 2L \Gamma }
\end{equation}
We will assume this optimal aperture to be adopted throughout the rest of this paper. 

\begin{figure}\centerline{{\includegraphics[width=10cm]{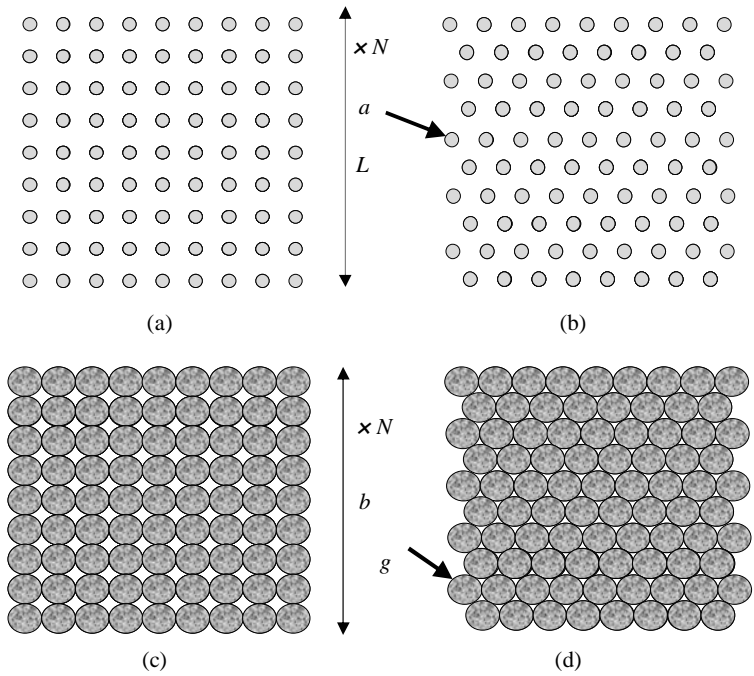}}}
  \caption{(a) and (b): grids of windows that let the laser light reach the object surface in (a) the rectangular and (b) the equilateral symmetries. (c) and (d): Speckle patterns at the CCD detector for (c) the rectangular and (d) the equilateral arrangements.}
  \end{figure}
  
The non-overlapping condition for the holes in the mask is
\begin{equation}
2aN\leq L
\end{equation}
which leads to
\begin{equation}
s \geq \frac{1.22 N \lambda f}{ L} = \frac{1.22  \lambda f}{ \Delta}
\end{equation}
or
\begin{equation}
s \geq  {1.22  \lambda  F }
\end{equation}
so that speckles in MWDSP are never smaller than in DDSP, provided that both systems have the same aperture.
 
\section{Comparison}

\begin{table}[h]
\caption{Results for DDSP and MWDSP}
\begin{center}
\begin{tabular}{c c c} \hline \hline 
Parameter & DDSP & MWDSP \\ [1ex] \hline 
Speckle size $s$ & $ \displaystyle 1.22 \lambda  F  $ & $ \displaystyle  \frac{1.22\lambda f}{2 a} $   \\ [1ex]
Range $\Gamma$ & $ \displaystyle \frac{b }{2L  F } $ & $ \displaystyle  \frac{b}{2 f N} $ \\ [1ex]
Lateral resolution $\Delta$ & $ \displaystyle \frac{f}{ F } $ & $ \displaystyle  \frac{L}{N} $
 \\ [1ex]
Quotient $\displaystyle \frac{\Gamma}{\Delta}$ & $ \displaystyle \frac{b}{2fL} $ & $ \displaystyle \frac{b}{2fL} $ \\ [1ex] \hline \hline
\end{tabular}
\end{center}
\end{table}

We are now going to compare the measuring range and lateral resolution of the DDSP and MWDSP systems. Table 1 summarises the results derived in the previous sections.

In DDSP there is the possibility to change the f-number $ F $ of the system. As for most popular cameras, $1\leq  F  \leq 20$, then we have
\begin{equation}
\Gamma_{DDSP} \leq \frac{b}{2L}
\end{equation}
and
\begin{equation}
0.05 f\leq \Delta_{DDSP} 
\end{equation}
so that the measuring range is only limited by the $\frac b L$ relation. 

In MWDSP, the aperture is related to $\Gamma$ through the equation
\begin{equation}
 F  = \frac{b}{2L \Gamma_{MWDSP}}
\end{equation}
and to the lateral resolution through
\begin{equation}
\Delta_{MWDSP} = \frac{f}{ F }
\end{equation}
so that the system aperture relates identically to the measuring parameters in DDSP and MWDSP. Thus
\begin{equation}
\Gamma_{MWDSP} \leq  \frac{b}{2L} 
\end{equation}
\begin{equation}
0.05 f\leq \Delta_{MWDSP} 
\end{equation}

Finally we arrive at the relation
\begin{equation}
\displaystyle \frac{\Gamma}{\Delta} = \displaystyle \frac{b}{2fL}
\end{equation}
which mutually limits the measuring range and lateral resolution of the system in both DDSP and MWDSP.

When considering the illumination power, it is straightforward to assert that if the aperture is the same in both systems, then there is more efficiency in DDSP than in MWDSP. In general, the relation of power at the detector is
\begin{equation}
\frac{I_{MWDSP}}{I_{DDSP}} = \frac{\pi N^2 a^2}{L^2}
\end{equation}
where $I_{DDSP}$ and $I_{MWDSP}$ are the mean intensities in the DDSP and MWDSP systems. Substituting according to the contents of Table 1, yields
\begin{equation}
\frac{I_{MWDSP}}{I_{DDSP}} = \frac{\pi a^2}{\Delta^2} =  \pi \displaystyle \frac{\left(\frac{1.22 \lambda f}{2 s_{MWDSP}}\Delta_{DDSP}\right)^2}{\left(\frac{1.22 \lambda f}{s_{DDSP}}\Delta_{MWDSP}\right)^2}
\end{equation}
and simplifying we get
\begin{equation}
\frac{I_{MWDSP}}{I_{DDSP}} = \frac{\pi}{4} \left( \frac{ s_{DDSP} \, \Delta_{DDSP}}{ s_{MWDSP} \,\Delta_{MWDSP}} \right)^2
\end{equation}
or
\begin{equation}
\frac{I_{MWDSP}}{I_{DDSP}} = \frac{\pi}{4} \left( \frac{ s_{DDSP} \, F _{MWDSP}}{ s_{MWDSP} \,  F _{DDSP}} \right)^2 
\end{equation}
which, on account of Eq.(26), gives
\begin{equation}
\frac{I_{MWDSP}}{I_{DDSP}} \leq \frac{\pi}{4}
\end{equation}
Finally, substituting for the apertures in Eq.(37) yields
\begin{equation}
\frac{I_{MWDSP}}{I_{DDSP}} = \frac{\pi}{4} \left( \frac{ s_{DDSP}\, \Gamma_{DDSP}}{ s_{MWDSP}\, \Gamma_{MWDSP}} \right)^2
\end{equation}
When both apertures and speckle sizes are equal, the MWDSP power efficiency is about $78.5\%$ that of DDSP. If equilateral triangular arrays are used, this result is slightly improved. However, under these conditions, the mean intensity within the speckle circles in the MWDSP system and $I_{DSP}$ are equal. In MWDSP we can increase the power at the detector by reducing either the measuring range (and accordingly the aperture and lateral resolution) or the speckle size or both, in order to get a similar intensity. Nevertheless, when the speckles in MWDSP are larger than in DDSP, the intensity in the latter is always greater than in the former. This is the main disadvantage of MWDSP that, in our opinion, is outweighed by the considerations that follow. 

If we now take into account that the resolution of the detector is $\delta$ there are more restraints to be fulfilled. Speckles need to be at least as wide as $\delta$ so that there is no averaging of speckles in the recorded picture. This poses a serious limitation on $DDSP$ since both its speckle size and measurement properties depend on $ F $; however, in MWDSP, speckle size only depends on $a$, so that it does not alter the choice range for $ F $ and keeps the flexibility of the system untouched. Moreover, there is no problem in reducing the value of $a$ (as long as we are far from $\lambda$) making speckles as big as desired. In fact the only restriction on MWDSP is that
\begin{equation}
2Na \leq L \Rightarrow s_{MWDSP}\geq 1.22\lambda  F 
\end{equation}
so that the $a$ parameter can always be adjusted to generate speckles whose size is $24.4 \lambda$ which is far from making speckles too large, since generally $\delta$ is not much smaller than $24.4\lambda$. In other words, for each value of $F$ there is a minimum size of speckle ($1.22 \lambda F$) , which is always attainable by choosing the maximum admissible $a$. Even for the maximum value of $F$, the minimum obtainable speckle size is not too large for the usual resolutions of CCD cameras.

This consideration also applies for DDSP, since 
\begin{equation}
s_{DDSP} = 1.22 \lambda  F 
\end{equation}
is in the $ F  = 20$ limit equal to $24.4 \lambda$ which is not much bigger than $\delta$. 

Consequently, in practice, speckle size considerations only restraint the measuring range of DDSP, leaving the lateral resolution limit around $0.05 f$. Besides, the speckle size in MWDSP can be freely raised from the DDSP value to several times $\delta$ by reducing the $a$ parameter, without changing the system aperture. This possibility makes that the DDSP limitation on the measuring range does not hold for MWDSP.

Table 2 lists the measuring characteristics for both systems when $\delta = s$.

\begin{table}[h]
\caption{Results for DDSP ($s=\delta$) and MWDSP}
\begin{center}
\begin{tabular}{c c c} \hline \hline 
Parameter & DDSP & MWDSP \\ [1ex] \hline 
Speckle size $s$ & $ \displaystyle \delta $ & $ \displaystyle  \frac{1.22 \lambda f}{2 a} $    \\ [1ex]
Range $\Gamma$ & $ \displaystyle \frac{1.22 b \lambda}{2L\delta} $ & $ \displaystyle  \frac{b}{2 f N} $ \\ [1ex]
Lateral resolution $\Delta$ & $ \displaystyle \frac{1.22 f \lambda}{\delta} $ & $ \displaystyle  \frac{L}{N} $ \\ [1ex]
Quotient $\displaystyle \frac{\Gamma}{\Delta}$ & $ \displaystyle \frac{b}{2fL} $ & $ \displaystyle \frac{b}{2fL} $  \\ [1ex] \hline \hline
\end{tabular}
\end{center}
\end{table}

Typical values for $\delta$ are in the vicinity of 10-20 microns, so that if $L/b\approx 5$, $\Gamma_{DDSP}$ is in the order of $10^{-3}$ radians for $\delta=20$ microns, whereas $\Gamma_{MWDSP}$ can be up to $b/(2L) \approx 10^{-1}$ radians. Lateral resolution can be reduced up to $0.05 f$. Using a $20mm$ lens, the minimum $\Delta$ is around $1mm$ for DDSP and MWDSP.

\section{Conclusion}

By suitably varying the optical aperture, different measuring ranges and lateral resolutions are attainable in the DDSP-system, almost exactly as by varying the $N$ parameter in the MWDSP-system. However, the optical aperture in the first system has severe limitations on account of speckle size considerations. On the other hand, in the MWDSP-system, speckle size is set by properly choosing the $a$ parameter independently from the selection of $N$. The main drawback in MWDSP is its slightly poorer intensity efficiency, which is outweighed by its benefits in those applications where the DDSP measuring range falls too short. MWDSP is especially well suited for use with bending plates, where large tilt ranges are needed. It could be mentioned, as yet another advantage of the MWDSP-system, its adaptive potential. Work on masks with non-equal spacing of holes is currently in progress, so that the measuring range and the lateral resolution can be changed across the object, achieving a fine lateral resolution in some points of limited slope change values and a coarser lateral resolution in points with greater expected slope changes. A second enhancement possibility is the utilisation of multiple masks (before and after the transformation) in order to increase lateral resolution and overcome the limiting relation (see Eq.(33)) that holds in both systems.

\end{document}